\documentstyle[12pt,aasms]{article}
\journalid{337}{15 January 1989}
\articleid{11}{14}
\begin{document}

\title{TGRS MEASUREMENTS OF THE POSITRON ANNIHILATION SPECTRUM\\
FROM THE GALACTIC CENTER}

\author{M. J. HARRIS\footnote{Universities Space Research Association,
harris@tgrs2.gsfc.nasa.gov}, B. J. TEEGARDEN, T. L. CLINE, N. GEHRELS, 
D. M. PALMER\altaffilmark{1}, R. RAMATY, AND H. SEIFERT\altaffilmark{1}}
\affil{Code 661, NASA/Goddard Spaceflight Center, Greenbelt, MD 20771}

\begin{abstract}

We have obtained spectra of the Galactic center at energies 400--600 keV 
from high-resolution data acquired by the TGRS Ge spectrometer on board 
the {\em WIND\/} mission during 1995--1997.  The data were obtained using
an on-board occulter, and are relatively free from systematics and 
backgrounds.  Analysis of the spectra reveals a well-resolved electron-positron
annihilation line at 511 keV and the associated continuum 
due to annihilation via positronium formation.  Measurements of the line
width and the continuum-to-line ratio allow some constraints to be placed
on the interstellar sites where annihilation occurs.

\end{abstract}

\clearpage

\section{Introduction}

The line at 511 keV from the annihilation of electrons and
positrons in the region of the Galactic center (GC) is the best-studied line
in $\gamma$-ray astronomy.  Over 20 years of observations (reviewed by
Tueller 1993) have established that there is an extensive diffuse
line source of total intensity $\sim 2 \times 10^{-3}$ photon cm$^{-2}$
s$^{-1}$.  This source has recently been mapped in considerable detail
by OSSE on board the {\em Compton Observatory\/} (Purcell et al. 1997), 
which has revealed a third spatial component
in addition to the well-known Galactic disk and bulge components.  This new
component is extended and is centered at $l = -2^{\circ}$, 
$b = +9^{\circ}$, well above the Galactic plane.  
It is unclear whether there are any
point sources superimposed on this diffuse distribution; recent results do
not show any variability in the flux.  The line is known to be narrow and 
centered at 511 keV (Leventhal, MacCallum, \& Stang 1978).  
The annihilation spectrum also includes
a lower-energy continuum arising from $3 \gamma$ annihilation via 
the formation of positronium (Ps).

In principle, spectral lines contain much information about the
physical conditions in the line formation region.  The next step in the 
study of the 511 keV line will be to extract the information
contained in the line profile and in the ratio of line to Ps 
continuum amplitudes.  The key requirement is for sensitive long-term 
measurements with fine spectral resolution.  The measurements described 
above were mostly made with low-resolution scintillator detectors.  In this 
paper, we describe observations made over more than 2 years with the 
high-resolution Ge spectrometer TGRS on board the {\em WIND\/} spacecraft.

\section{Observations and Analysis}

The Transient Gamma-Ray Spectrometer (TGRS) is mounted on the top (+$z$)
surface of the {\em WIND\/} spacecraft.   Since launch in 1994 November 
{\em WIND\/} has been in a complicated extremely elliptical Earth orbit,
which has avoided background $\gamma$-radiation due to factors such as
Earth albedo and particle irradiation in the radiation belts.  Throughout
this time the top surface has pointed towards the South Ecliptic Pole,
and the spacecraft has rotated with a period of 3 s.

The TGRS detector itself is a radiatively cooled 35 cm$^{2}$ Ge crystal
sensitive to energies between 20 keV--8 MeV.  The instrument gain
has varied slightly (by $\sim 0.5$\%) during the mission; this
variation has been monitored and corrected for.  There has
been some loss of resolution during the mission caused
by radiation damage due to cosmic-ray impacts
on the detector.  The detector is unshielded, and 
therefore monitors the whole +$z$ hemisphere for $\gamma$-ray bursts in 
the above energy range (Palmer et al. 1996).  However, the 
results in this paper are obtained from a different data set.  The 
rotation of the spacecraft offers the opportunity of modulating the signal 
from a constant source, which has been realized by adding a narrow 
1-cm thick lead occulter to the instrument.  This occulter is 
concentric with the detector and subtends $90^{\circ}$ as seen from it.  
It is almost in the rotation plane\footnote{ The occulter is set slightly 
above (+$z$) the detector, by about 5.5$^{\circ}$, so that instead of 
tracking the ecliptic plane, it is offset 5.5$^{\circ}$ south of it.  It 
therefore occults the Galactic center.  The breadth of the occulter is 
$\simeq 16^{\circ}$ FWHM as seen from the detector.}.  The spatial arrangement
and sky coverage of the occulter are illustrated by Teegarden et al.
(1996).

The occulted data are transmitted from the spacecraft for energies
20 keV--1 MeV in four broad energy windows, each binned into 64
channels.  Only one window is binned at the best possible resolution
for a Ge detector ($\simeq 1$ keV per channel); this window is
centered on 511 keV and provides fine detail over the region
479--543 keV.  The data in all windows are further binned into 128
angular channels (``sectors" of $2.8125^{\circ}$ in ecliptic longitude).

The effect of modulation by the occulter on the count rate in the
high-resolution window during a 2
year period is shown in Fig. 1.  The two cosmic sources which are clearly
occulted, at ecliptic longitudes $\lambda \simeq 84^{\circ}$ and
$267^{\circ}$, are the Crab and the GC
respectively.  Note that the
precise shapes of the occultation ingress and egress
contain information about the extension of the sources along the
ecliptic plane; for example in Fig. 1 the GC source is not well
described by the detector's response to an occulted point source, but
seems to be extended, as discussed below. 

The amplitude of the occultation dip in Fig 1. is a direct measure of
the intensity of the source.  When it is measured for each energy channel
a spectrum of the GC is obtained (Fig. 2).  As implied by Fig. 1, these
amplitudes are obtained by fitting the occulted count rate with a model
of the detector's response to a diffuse source\footnote{ The Crab spectrum 
obtained by fitting a point source to the other occultation dip is in
good agreement with recent measurements (Bartlett 1994).}.  We took into account
the systematic error due to the uncertainty in the GC source distribution,
which is visible in Fig. 1 as the difference in the amplitude which would
be derived from the two responses, one of which is spatially extended.

Slight modifications (described below in each case) were made in the analyses 
by which the basic data in Figs. 1 and 2 were obtained in order to derive the
annihilation-related physical quantities.  These are the intensity, width
and energy of the narrow 511 keV line, the intensity 
of the Ps continuum (visible as a step at 511 keV in Fig. 2), the position 
of the line source, and its spatial extent.

{\em (1) Line intensity.\/}
Although the spectrum around 511 keV is dominated by the annihilation
line (Fig. 2), its intensity will be systematically overestimated if the
underlying continuum is neglected.  We therefore fitted
this region of the spectrum by a model containing three 
components, the line itself (a Gaussian with three free parameters, 
intensity, width, and centroid), the Ps continuum, and an underlying power law.
The fit was performed by folding the models through the instrument
response function\footnote{ The instrument response was determined by
Monte Carlo simulations of the TGRS-{\em WIND\/} assembly using the
standard GEANT code (Seifert et al. 1997).} 
and varying the parameters of the components until the
best agreement was obtained with the count spectrum (as determined by the
minimum of $\chi^{2}$).  The uncertainty in 
each parameter was obtained from the region over which $\chi^{2}$ exceeded
this minimum value by 1, after re-minimizing with all other parameters allowed 
to vary (Lampton, Margon \& Bowyer 1976, Avni 1976).  To monitor the
{\em variability} of the line, the period January 1 1995--March 13 1997 was
divided into nine 90 d intervals for which spectra were extracted and
fitted in the same way.  Effects of nonlinearity when these fits are
combined were regarded as a systematic error in the overall intensity.

{\em (2) Line Width.\/}
A direct measurement of the line width from the spectrum of Fig. 2 would
not take into account the long-term deterioration of instrument resolution.  
We therefore measured the line widths from the 90 d
spectra described above.  These measured total widths $\sigma_{tot}$ are
resultants of the cosmic line widths $\sigma_{gal}$ and the instrumental
width $\sigma_{inst}$ added in quadrature,
$\sigma_{tot}^{2} = \sigma_{inst}^{2} + \sigma_{gal}^{2}$.  To establish
$\sigma_{inst}$, we used two strong lines on either side of 511 keV which
arise in the detector and appear in TGRS background spectra.  These lines,
from $^{69}$Zn (439 keV) and $^{69}$Ge (584 keV ), are
routinely monitored to determine instrument gain shifts.  Their intrinsic
widths are very narrow so that their measured widths reflect 
$\sigma_{inst}$ at the appropriate energies.  We measured these widths at
90 d intervals and interpolated to determine $\sigma_{inst}$ at the position
of the 511 keV line.

The nonlinear relation between $\sigma_{gal}$ and $\sigma_{tot}$ introduces
a correlation between $\sigma_{gal}$ and the error obtained by
differentiating that relation.  Wheaton et al. (1995) showed that 
when an average is taken with weights derived from the errors, such a
correlation may cause a systematic error in the final value, in the sense
that $\sigma_{gal}$ would tend to be overestimated.  We avoided this
problem by combining the individual 90--d $\sigma_{gal}$ values unweighted,
which gives a smaller average.  The difference between weighted and
unweighted results is regarded as a possible systematic error.

{\em (3) Line Energy.\/}
Gain correction of the TGRS instrument is
performed by measuring the energy channels corresponding to lines in the 
background spectrum at known energies.  Below 1
MeV the channel shifts are almost, but not exactly, proportional to the line
energy.  There is enough scatter about this linear relationship that using 
the average gain shift per keV
from several lines would introduce a small systematic error ($\sim 0.3$ 
keV at energy 511 keV).

In the present case we consider only energies which are 
close to 511 keV, near which there is a strong background line due to positron
annihilation in the detector.  If we assume that, over our small energy 
range, the gain shift per keV is constant, and that the physical energy 
of the background line is exactly 511.00 keV, we can avoid
the systematic error by using this line alone for gain correction.  We
therefore measured the background line energy, corrected it to 511.00
keV, and measured the cosmic line energy relative to it.  There is 
a small systematic error in the measurement of
the background line energy (the statistical error is negligible)
due to small deviations of the line shape 
from Gaussian and to uncertainties in the continuum, whose effects were
estimated by altering these parameters in our fitting of both cosmic and 
background spectra.

{\em (4) Positronium Continuum.}
Only a small part of the full continuum is visible in Fig. 2.  To
improve the statistics of our fit, we obtained and fitted spectra,
by the same method as illustrated in Figs. 1 and 2, for energies down to 
400 keV where the continuum is dominated by Ps, and also up to 1 MeV to
include the underlying Galactic continuum.  We checked for systematic 
errors in this fit due to the shape of the underlying Galactic continuum,
by repeating it using thermal bremsstrahlung, inverse Compton and synchrotron 
spectra instead of a power law.  No significant change in either $\chi^{2}$
or the fit parameters was found.

{\em (5) Source position and extension}
It is by no means obvious that the three spectral components in Fig. 2
all share the same spatial distribution.  Rather than fit the occultation
response function to data in which all three are combined, as in Fig. 1,
we attempted to isolate the narrow 511 keV line before fitting.  This 
was done by accumulating the occulted data over a very narrow window
506--516 keV which is dominated by the line.

The detector occultation response was fitted to these data assuming a
prescribed source distribution, i.e. a Gaussian of variable width
along the ecliptic plane.  Extensive analyses of the response are necessary
before the next step can be taken, in which actual properties of 
the source distribution are deduced.  These analyses are 
in progress; here we present preliminary results obtained using the
Gaussian hypothesis.  We therefore make no estimate of systematic errors
due to the unknown source distribution.  Using data such as those of Fig. 1
we can however estimate errors due to contamination by the other spectral
components.

\section{Results, Discussion and Conclusions}

The results of our measurements of the annihilation spectrum are given in 
Table 1.  These results supersede the preliminary measurement made by Teegarden
et al. (1995), which reported a 511 keV line flux $1.64 \times 10^{-3}$
photon cm$^{-2}$ s$^{-1}$.  Two features of the earlier analysis contributed
to this overestimate.  First, the line flux was obtained from the count rate in
the narrow 506--516 keV band, without fitting the shape of the spectrum,
and is thus overestimated by including the other spectrum
components (Fig. 1).  Second, the present analysis uses an improved model
of the instrument spectral response, instead of simply dividing
by photopeak effective area as was done in the earlier work.

\subsection{Comparison with OSSE results}

We calculated the flux, dimension and centroid of the OSSE model of 
the GC 511 keV emission (Purcell et al. 1997) as folded through the TGRS 
occulted response (Table 1).  Since the occulter passes close to the centers
of two of the three spatial components of the model (exactly crossing
the GC, and $2.8^{\circ}$ from the high-latitude feature) a test of these
model features becomes possible in principle.  
The measurements of the flux and centroid are in good agreement;
the offset of the TGRS centroid
measurement from the GC is in the same direction as the offset of the
OSSE centroid due to the new high-latitude feature, but is also
compatible with the GC.  The spatial extension found by TGRS slightly exceeds
that found by OSSE, but to draw any conclusion from 
this would be premature since improved modeling of the occultation response
is required.  Our results agree with OSSE in finding a lack of 
variability on 90 d time-scales (Fig. 3).

\subsection{Source physics: Line width}

Our measurements of the total line width $\sigma_{tot}$, and of the
background line widths, are shown in Fig. 4.  The interpolated instrument
intrinsic width $\sigma_{inst}$ is also shown.  It is clear that 
$\sigma_{tot}$ at all times exceeds $\sigma_{inst}$; this is necessary
if the cosmic line width is to be obtained from
$\sigma_{tot}^{2} = \sigma_{inst}^{2} + \sigma_{gal}^{2}$.  The result
(Table 1) is somewhat narrower than the average of four balloon
measurements by the GRIS Ge detector (Leventhal et al. 1993), but the
difference is not very significant.

The line width reflects the convolved widths of components due to
different annihilation mechanisms predominating in different phases
of the ISM.  These mechanisms were treated by Guessoum, Ramaty \&
Lingenfelter (1991).  They may be divided into two classes.  Firstly,
annihilation by charge-exchange in flight produces a broad line (FWHM
6.4 keV), and is predominant in cold molecular clouds.  The second
class contains all other processes, which produce lines narrower than the
instrument resolution.  We can therefore hope to test two alternative
suggestions by Guessoum et al. --- annihilation occurring uniformly in
all phases of the ISM, and otherwise-uniform annihilation excluding cold
clouds.

We therefore repeated our analysis under the assumption that the width
$\sigma_{gal}$ had two components, a broad component of width
6.4 keV, and a narrow unresolved component.  Instead of line
width, we now have the amplitude of the 6.4-keV broad component as a
fitted parameter.  The spectra were fitted equally well by this model;
though there were no significant
improvements in the $\chi^2$ values, we hope this procedure yields
physical insight into the meaning of $\sigma_{gal}$.

Assuming the presence of a 6.4-keV broad line component, we found that
$11$\%$\pm 9$\% of the total line intensity was due to this broad line.
This is much closer to the prediction when 
positrons are excluded from molecular cloud cores (in which case the broad
line contributes only 11\%: Guessoum et al. 1991) than
to the maximum predicted broad-line contribution of 59\% when positrons 
penetrate all phases of the ISM equally.

\subsection{Source physics: Positronium fraction}

The fraction $f$ of positrons which annihilate through the formation of Ps
can be written $f = 2/[2.25(I_{511}/I_{Ps})+1.5]$,
where $I_{511}$ and $I_{Ps}$ are the line and Ps continuum intensities
(Brown \& Leventhal 1989); our result from Table 1 is
$f = 0.94 \pm 0.04$.\footnote{ The uncertainty does not include those systematic
errors in $I_{511}$ and $I_{Ps}$ in Table 1 which are positively
correlated.}  This is in good agreement with the most recent OSSE
result $f = 0.97 \pm 0.03$ (Kinzer et al. 1996).  However, predicted values
from annihilation in most of the phases of the ISM cluster in the
range $f \sim 0.9$--1.0, so small discrepancies in 
measured $f$ may be important.

Our result falls roughly in the middle of this range, and is consistent
with annihilation in cold molecular clouds ($f = 0.9$: Brown, Leventhal
\& Mills 1986),
the warm neutral or ionized ISM ($f = 0.9$--0.95: Bussard, Ramaty \&
Drachman 1979), or any combination of these (Guessoum et al. 
1991).  It is not compatible with annihilation in the hot phase, nor
with any scenario in which grains are important sites of annihilation.
These two statements are in fact equivalent, since in the hot phase
grains become the most important location for annihilation in the
absence of H atoms.  The corresponding value of $f$ is expected to be
very low ($\le 0.5$: Guessoum et al. 1991).

\subsection{Summary}

We have measured the 511 keV line from the GC and also the Ps continuum
associated with it during 1995--1997.  Our values for the intensities
of these features agree with the most recent OSSE measurements.  Our
preliminary results for the spatial distribution of the line are
consistent with the OSSE mapping, but require further analysis of the
instrument response.  The 511 keV line is resolved, and, if a specific
model for its width is assumed (an underlying broad component from
annihilation through charge-exchange in flight), then our result favors
a scenario in which annihilation in cold molecular clouds is
suppressed.  Our measurement of the Ps fraction $f$ from the Ps 
continuum is consistent with this, and suggests further that annihilation
in the hot phase of the ISM is of minor importance.

\acknowledgments

We are grateful to Theresa Sheets (LHEA) and Sandhya Bansal (HSTX) for
assistance with the analysis software, and to the referee for many
detailed suggestions.

\clearpage

\begin{figure}

\caption{Count rate in the energy range 479--543 keV, accumulated by 
sector bin between 1995 January 1 and 1997 March 13.  Dashed line --- fitted
TGRS response to a point source.  Full line --- fitted response to a 
Gaussian distribution along the ecliptic centered on the GC with FWHM 
$30^{\circ}$.  Arrows --- ecliptic longitudes of the Crab and the GC.}

\caption{Count spectrum obtained from channel-by-channel analysis of the data
in Fig. 1 in the TGRS high-resolution energy window, assuming Gaussian
distributed source of FWHM $30^{\circ}$.  The spectrum is
fitted with three components: dotted line --- line at 511 keV;
dot-dashed line --- positronium continuum; dashed line --- power law
continuum.}

\caption{Fluxes measured in the 511 keV line at 90 d intervals from 1995
January 1.}

\caption{Widths of the 511 keV line and the background calibration lines 
measured at 90 d intervals.  Full lines --- widths of the 584 keV 
background line (top) and the 439 keV line (bottom).  The uncertainties
in these widths are too small to be drawn to scale.  Dashed line ---
interpolated instrumental width $\sigma_{inst}$ at 511 keV.  Data points
--- measured width of the 511 keV line $\sigma_{tot}$.}

\end{figure}

\clearpage

\begin{table*}
\begin{center}
\begin{tabular}{lccl}
\tableline
Quantity & Measured value~\tablenotemark{a} & Previous values & Units \\
\tableline
511 keV line amplitude & $1.07 \pm 0.05^{+0.05}_{-0.08} \times 10^{-3}$ & 
$1.25 \pm 0.38 \times 10^{-3}~\tablenotemark{b}$ & photon cm$^{-2}$ s$^{-1}$ \\
Line width (FWHM) & $1.81 \pm 0.54 \pm 0.14$ & $2.5 
\pm 0.4$~\tablenotemark{c} & keV \\
Line energy~\tablenotemark{d} & $510.98 \pm 0.10 \pm 0.04$ & 
$510.92 \pm 0.23$~\tablenotemark{e} & keV \\
Ps continuum amplitude & $3.8 \pm 0.6^{+0.5}_{-0.6} \times 10^{-3}$ & \ldots & 
photon cm$^{-2}$ s$^{-1}$ \\
Source dimension (FWHM)\tablenotemark{f} & $24^{\circ} \pm 5^{\circ}
~^{+3^{\circ}}_{-4^{\circ}}$ & $14^{\circ}$~\tablenotemark{b} & \ldots \\
Source centroid~\tablenotemark{f} & $265.5^{\circ} \pm 1.2^{\circ} \pm 
1.0^{\circ}$ & 
$264.8^{\circ}$~\tablenotemark{b} & ecliptic longitude 
$\lambda$~\tablenotemark{g} \\
\tableline

\end{tabular}
\end{center}

\tablenotetext{a}{ Statistical errors given first, followed by
systematic errors.}
\tablenotetext{b}{ Calculated from OSSE model (Purcell et al. 1997).}
\tablenotetext{c}{ Leventhal et al. 1993.}
\tablenotetext{d}{ Relative to background line at 511.00 keV.}
\tablenotetext{e}{ Mahoney, Ling \& Wheaton 1994.}
\tablenotetext{f}{ Assuming a Gaussian distribution along the ecliptic.}
\tablenotetext{g}{ The GC is at $\lambda=266.8^{\circ}$.}

\caption{Results from analysis of TGRS measurements 1995--1997.}

\end{table*}

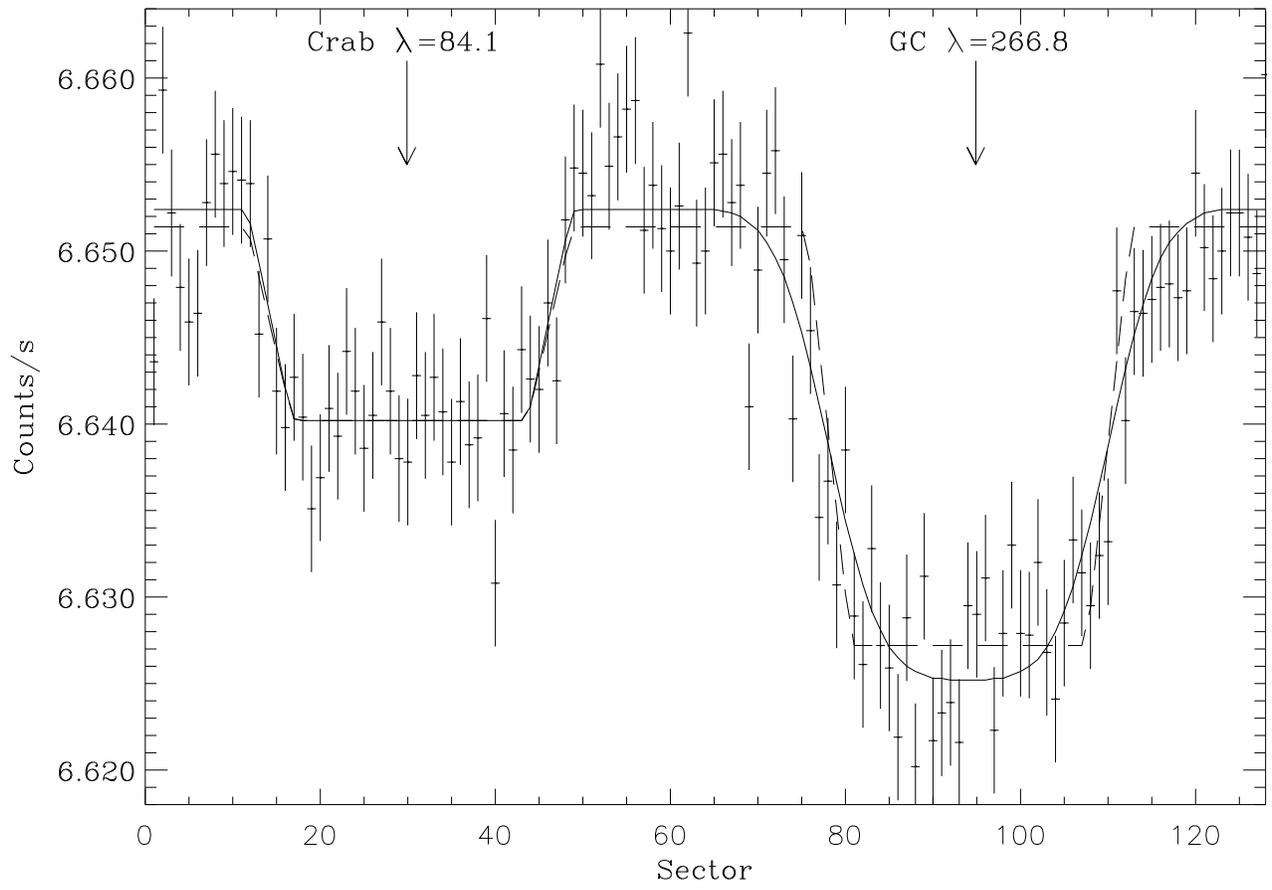
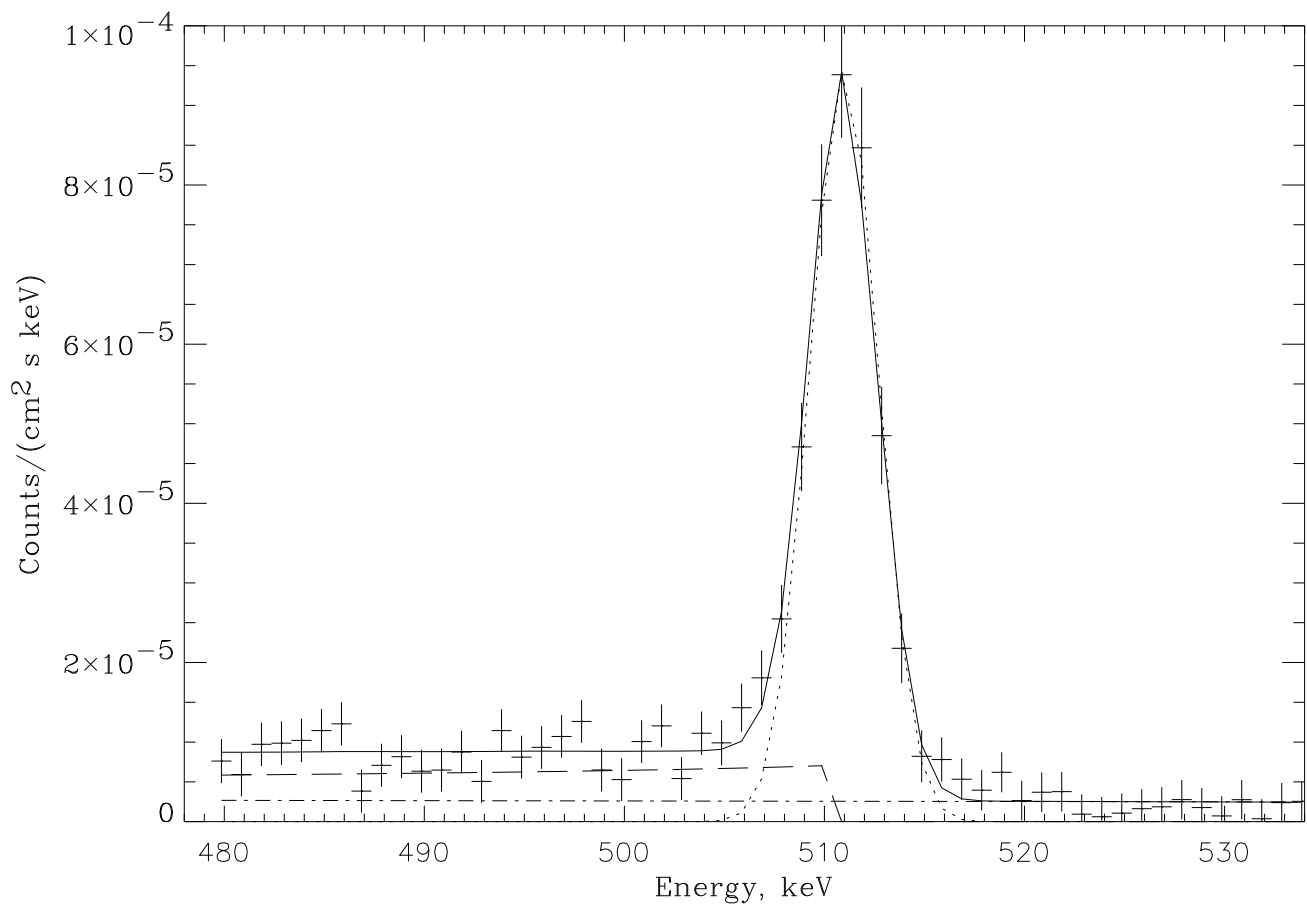
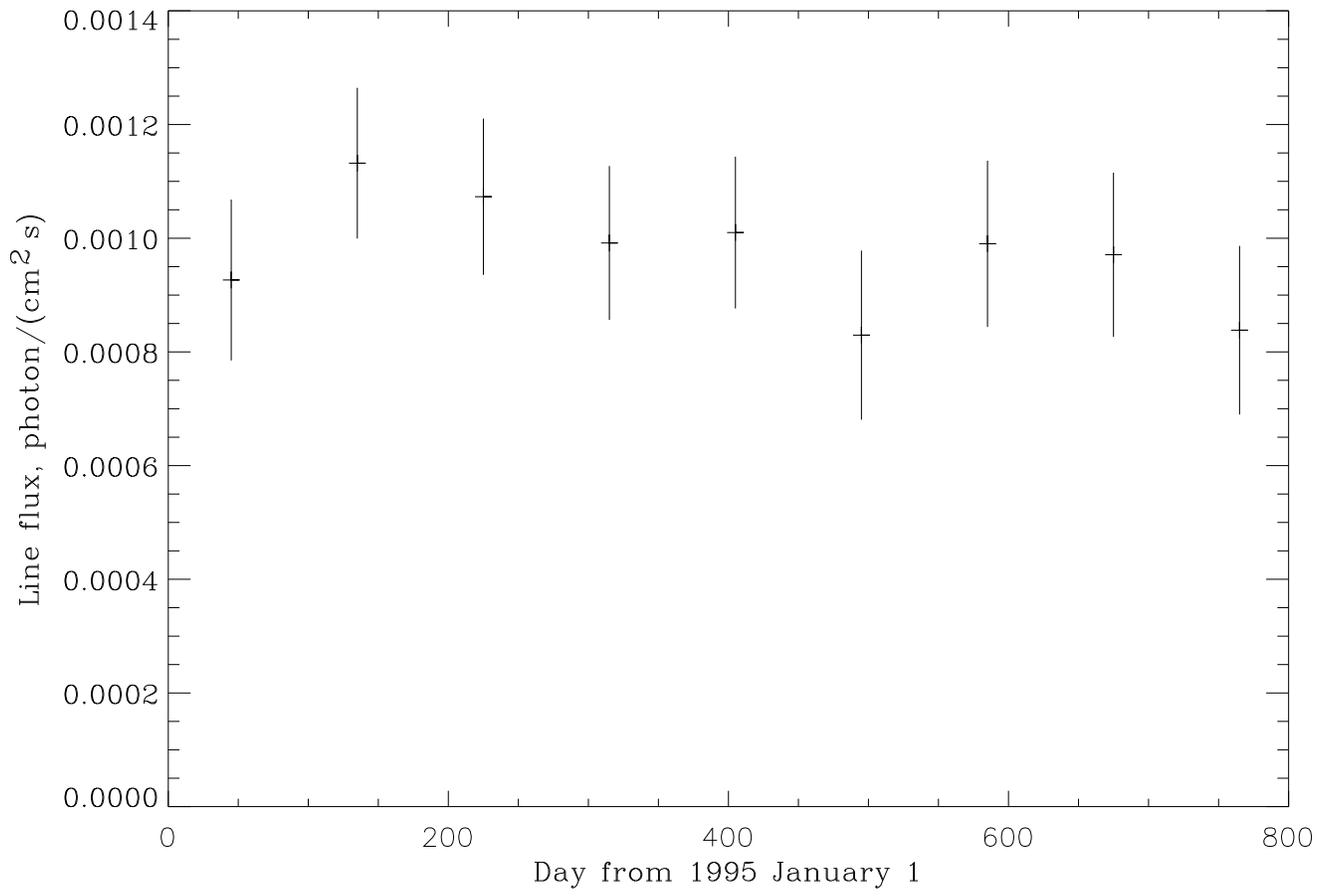
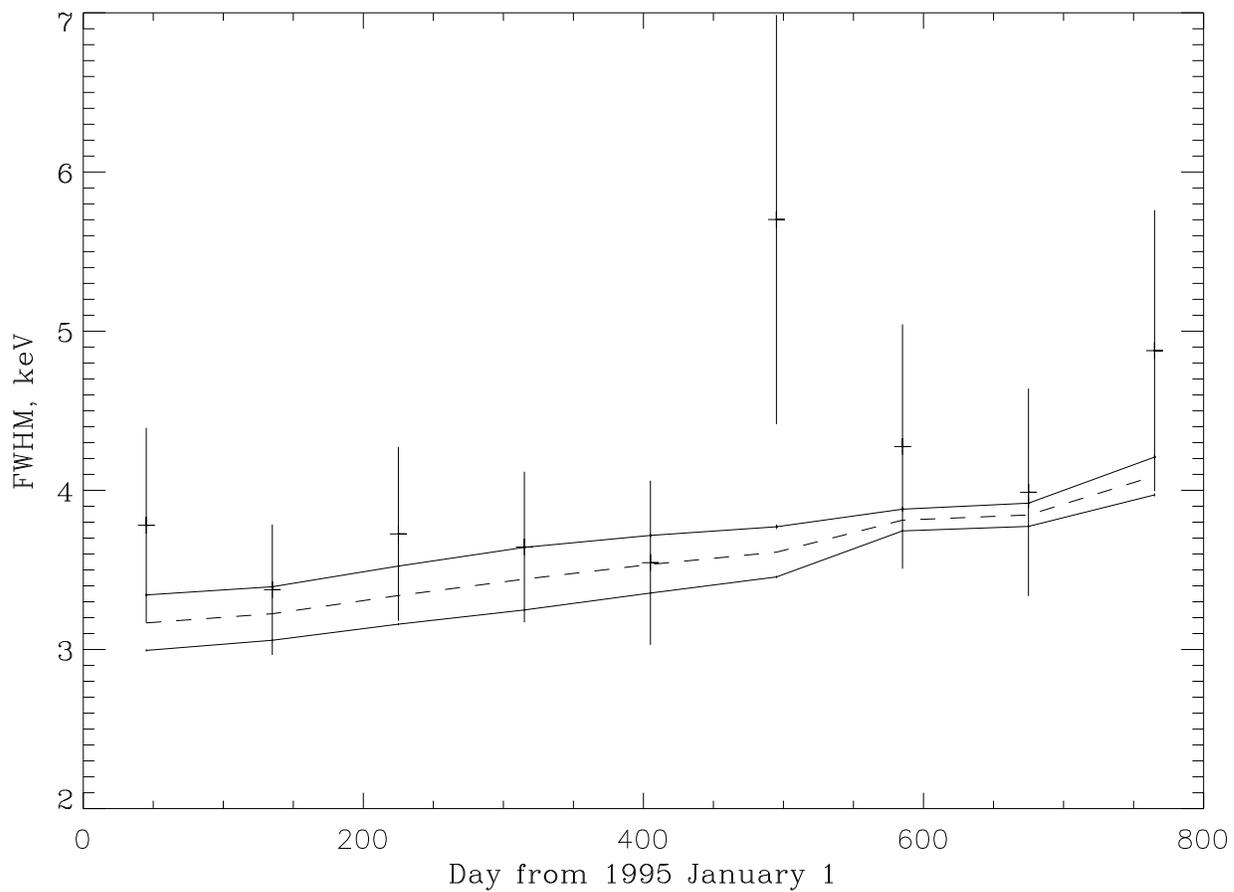
\end{document}